\documentstyle[12pt]{article}
\parskip=\medskipamount

\title{On zero modes of the eleven dimensional superstring}
\author{A.A. Deriglazov\thanks{deriglazov@phtd.tpu.edu.ru}\\
Department of Mathematical Physics, Tomsk Polytechnical University,\\
634004 Tomsk, Russia,}
\date{and\\
D.M. Gitman\thanks{gitman@fma.if.usp.br}\\
Instituto de Fisica, Universidade de S\~ao Paulo,\\
P.O. Box 66318, 05315-970, S\~ao Paulo, SP, Brazil.}

\begin{document}
\maketitle
\large
\begin{abstract}
It is shown that recently pointed out by Berkovits on-shell degrees of
freedom of the $D=11$ superstring do not make contributions into
the quantum states spectrum of the theory. As a concequence, the
spectrum coincides with that of the $D=10$ type IIA superstring.
\end{abstract}
\noindent

In the recent work [1] the Green-Schwarz type formulation for the eleven
dimensional superstring action has been proposed. The action is invariant
under local fermionic $\kappa$-symmetry which eliminates half
of $\theta$-variables as well as under a number of global symmetries
which can be considered as a realization of the "new supersymmetry"
$S$-algebra [2--5]. It was also motivated that this $D=11$ model is
equivalent to the type IIA Green-Schwarz superstring.

In addition to coordinates of the $D=11$ superspace
$x^\mu(\sigma^a)$ , $\theta^\alpha(\sigma^a)$,
the action involves the auxilliary bosonic variables
$n^\mu(\sigma^a)$, $A^\mu_a(\sigma^b)$, which are designed to provide
the desired properties for the theory. A problem with these variables
was formulated in Ref. [6], where it was pointed out that their zero modes
can not be eliminated by means of gauge invariance and, hence, survive
in the sector of physical degrees of freedom \footnote{We are gratefull
also to J. Gates for bringing this fact to our attention}.
Thus, the  question of the
equivalence of this model to the GS superstring arises [6].
Since states spectrum of a string is determined by the action on a vacuum
of oscillator modes only, one expects that the presence of
zero modes for the case is inessential. In this short note we analyse
this problem more detaily in the canonical quantization framework.
It will be shown that taking into account of these zero modes do not
spoil the final conclusion of Ref. [1]. Namely, quantum states
spectrum of the model proposed coincides with that of the type IIA
superstring.

Since we are dealing with zero modes of even variables, let us first
consider a bosonic part of the action (we use the notations from Ref.1)
\begin{eqnarray}
S=\int d^2\sigma\left\{\frac{-g^{ab}}{2\sqrt{-g}}\partial_a x^\mu
\partial_b x^\mu-\varepsilon^{ab}\xi_a(n^\mu\partial_b x^\mu)
-n^\mu\varepsilon^{ab}\partial_a A^\mu_b-\frac 1\phi (n^2+1)\right\}.
\end{eqnarray}
By direct application of the Dirac-Bergmann algorithm one finds the
Hamiltonian
\begin{eqnarray}
H=\displaystyle\int d\sigma\left\{-\frac N2[\hat p^2+(\partial_1 x)^2]
-N_1(\hat p\partial_1 x)-\xi_0(n\partial_1 x)+(n\partial_1 A_0)+\right. \cr
\left.+\displaystyle\frac 1\phi (n^2+1)+\omega^{ab}(\pi_g)_{ab}+
\lambda_\phi\pi_\phi+\lambda_{\xi a}{\pi_\xi}^a+\lambda_0 p_0+
\lambda_1(p_1-n)+\lambda_n p_n\right\},\label{ham}
\end{eqnarray}
where it was denoted
\begin{equation}
\hat p^\mu\equiv\ p^\mu+\xi_1 n^\mu, \qquad N\equiv\frac{\sqrt{-g}}{g^{00}},
 \qquad N_1\equiv\frac{g^{01}}{g^{00}},
\end{equation}
and $p^\mu$, $p^\mu_a$, $p^\mu_n$, $(\pi_g)^{ab}$, $\pi^a_\xi$, $\pi_\phi$
are momenta conjugate to the variables $x^\mu$, $A^\mu_a$, $n^\mu$, $g^{ab}$,
$\xi_a$, $\phi$ respectively; $\lambda_*$ are Lagrange multipliers
corresponding to the primary constraints.
The full system of constrains can be presented as follows
\begin{equation}
 \qquad p^\mu_n=0, \qquad n^\mu-p^\mu_1=0;
\end{equation}
\begin{equation}
 \qquad \pi^1_\xi=0, \qquad \xi_1-(p_1 p)=0;
\end{equation}
\begin{equation}
 \qquad (\pi_g)_{ab}=0, \qquad \pi_\phi=0, \qquad \pi^0_\xi=0,
\qquad p^\mu_0=0;
\end{equation}
\begin{equation}
 \qquad (p_1)^2=-1, \qquad \partial_1 p^\mu_1=0;
\end{equation}
\begin{equation}
 \qquad H_0\equiv(p_1\partial_1 x)=0, \qquad H_\pm\equiv(\hat p^\mu\pm
\partial_1 x^\mu)^2=0;
\end{equation}
where the constraint $n^\mu=p^\mu_1$ was used. Constraints (4),(5)
are separated from others and form a system of second class,
while the remaining ones are first class. An appropriate gauge for
the constraints from Eq.(6) is
\begin{equation}
 \qquad g^{ab}=\eta^{ab}, \qquad \phi=2, \qquad \xi_0=0,
 \qquad A^\mu_0=\displaystyle\int\limits_0^\sigma d\sigma'\xi_1\hat p^\mu.
\end{equation}
This choice simplifies the subsequent analysis of the
$(A^\mu_1, p^\mu_1)$-sector since Hamiltonian equations of motion
for these variables look now as
\begin{equation}
 \qquad \partial_0 A^\mu_1=p^\mu_1, \qquad \partial_0 p^\mu_1=0.
\end{equation}

In order to find a correct gauge for the constraints (7) let us
consider Fourier decomposition of periodical in the interval
$\sigma\subset[0,\pi]$ functions
\begin{equation}
\begin{array}{l}
 \qquad A^\mu_1(\tau,\sigma)=Y^\mu(\tau)+\sum\limits_{n\ne0} y^\mu_n(\tau)
e^{i2n\sigma},\\
 \qquad p^\mu_1(\tau,\sigma)=P^\mu_y(\tau)+\sum\limits_{n\ne0}
p^\mu_n(\tau) e^{i2n\sigma}.
\end{array}
\end{equation}
Then the constraint $\partial_1 p^\mu_1=0$ is equivalent to $p^\mu_n=0$,
$n\ne0$, and an appropriate gauge is $y^\mu_n=0$, or, in the covariant
form $\partial_1 A^\mu_1=0$. Thus, physical degrees of freedom in the
sector $(A^\mu_1, p^\mu_1)$ are zero modes of these variables, and
the corresponding dynamics is
\begin{equation}
\begin{array}{l}
 \qquad A^\mu_1(\tau,\sigma)=Y^\mu+P^\mu_y\tau,\\
 \qquad p^\mu_1(\tau,\sigma)=P^\mu_y=const, \qquad (P_y)^2=-1.
\end{array}
\end{equation}
This sector of the theory (1) can be considered as describing a
string-like object $A^\mu_1(\tau,\sigma)$ which propagates without
oscillations with the center of mass $Y^\mu+P^\mu_y\tau$ and the
corresponding momenta $P^\mu_y$. The only quantum state of the
string is its ground state $\mid P_y>$ with mass $m^2_y=P^2_y=-1$.

Dynamics of the remaining variables is governed now by
\begin{equation}
 \qquad \partial_0 x^\mu=-p^\mu-(P_y p)P^\mu_y,
 \qquad\partial_0p^\mu=-\partial_1\partial_1 x^\mu.
\end{equation}
In addition, the constraints
\begin{equation}
 \qquad H_0\equiv(P_y\partial_1 x)=0,
 \qquad H_\pm\equiv(p^\mu+(P_y p)P^\mu_y\pm\partial_1 x^\mu)^2=0,
\end{equation}
hold, which obey the following algebra
\begin{equation}
\begin{array}{l}
 \qquad \left\{H_\pm,H_\pm\right\}=\pm4[H_\pm(\sigma)
\pm(P_y p)H_0(\sigma)+(\sigma\to\sigma')]
 \partial_\sigma \delta(\sigma-\sigma'),\\
 \qquad \left\{H_+,H_-\right\}=4[(P_y p)H_0(\sigma)+(\sigma\to\sigma')]
\partial_\sigma\delta(\sigma-\sigma'),\\
 \qquad \left\{H_0,H_\pm\right\}=\pm2 H_0(\sigma') \partial_\sigma
 \delta(\sigma-\sigma').
\end{array}
\end{equation}
On the $D=10$ hyperplane extracted by the constraint $H_0(\sigma)=0$ it
reduces to the standard Virasoro algebra. Note also that variable
$x^\mu(\tau,\sigma)$ obeys the free equation
$(\partial^2_\tau-\partial^2_\sigma) x^\mu=0$,
as a consequence of Eqs.(13),(14).

To proceed further, it is usefull to impose the gauge
\begin{equation}
(P_y\partial_1 p)=0,
\end{equation}
to the constraint $H_0=0$. By virtue of Eqs.(13),(16) one finds,
in particular, that $(P_y p)=(P_y P)$, where $P^\mu$ is the zero mode
of $p^\mu(\tau,\sigma)$. Then the final solution to Eq.(13) for the
case of closed world sheet reads
\begin{equation}
\begin{array}{l}
 \qquad x^\mu(\tau,\sigma)=X^\mu-
\frac 1{\pi}(P^\mu+(P_y P)P^\mu_y)\tau+\\ \qquad \qquad \qquad
\frac i{2\sqrt\pi}\sum\frac 1n[\bar\alpha^\mu_n e^{i2n(\tau+\sigma)}
+\alpha^\mu_{-n} e^{-i2n(\tau-\sigma)}],\\
\qquad p^\mu(\tau,\sigma)=\frac 1{\pi} P^\mu+
\frac 1{\sqrt\pi}\sum[\bar\alpha^\mu_n
e^{i2n(\tau+\sigma)}-\alpha^\mu_{-n} e^{-i2n(\tau-\sigma)}],
\end{array}
\end{equation}
which is accompanied by the constraints
\begin{equation}
 \qquad P^\mu_y\bar\alpha^\mu_n=0, \qquad P^\mu_y\alpha^\mu_{-n}=0,
\end{equation}
\begin{equation}
\begin{array}{l}
 \qquad H_+=\frac{8}{\pi}\sum\limits^\infty_{-\infty}L_n
e^{i2n(\tau-\sigma)},
 \qquad L_n\equiv\frac12\sum\limits^\infty_{-\infty}
\alpha^\mu_{n-k} \alpha^\mu_k=0,\\
 \qquad H_-=\frac{8}{\pi}\sum\limits^\infty_{-\infty}\bar L_n
e^{i2n(\tau+\sigma)},
 \qquad \bar L_n\equiv\frac12\sum\limits^\infty_{-\infty}\bar
\alpha^\mu_{n-k}\bar\alpha^\mu_k=0,
\end{array}
\end{equation}
where $\alpha^\mu_0=-\bar\alpha^{-\mu}_0\equiv\frac{1}{2\sqrt{\pi}}(P^\mu
+(P_y P)P^\mu_y)$.

From Eq.(18)and the equality $(P^\mu+(P_y P)P^\mu_y)P^\mu_y=0$ for
momenta of center of mass, it follows that the sector $(x^\mu,p^\mu)$
of the theory (1) describes in fact a closed string which lives on
the (D-1)-dimensional hyperplane ortogonal to the $P^\mu_y$ - direction.

From the zero modes $X^\mu$, $P^\mu$, $Y^\mu$, $P^\mu_y$, one
can construct the following quantities
\begin{equation}
 \qquad {\cal X}^\mu\equiv X^\mu-\frac12\frac{P_yY}{P_yP} P^\mu_y,
 \qquad {\cal P}^\mu\equiv P^\mu+(P_y P)P^\mu_y,
\end{equation}
with the properties
\begin{equation}
 \qquad \left\{{\cal X}^\mu,{\cal P}^\nu\right\}=\eta^{\mu\nu},
 \qquad \left\{{\cal X}^\mu,{\cal X}^\nu\right\}=\left\{{\cal P}^\mu,
{\cal P}^\nu\right\}=0.
\end{equation}
So the quantities ${\cal P}^\mu$, ${\cal L}^{\mu \nu}=\frac12 ({\cal X}^\mu
{\cal P}^\nu-{\cal X}^\nu{\cal P}^\mu)$ are generators of the
Poincare group.
This allows one to obtain the mass formulae for physical states. We adopt
the Gupta-Bleuler prescription by requiring that physical states be
annihilated by half of $:L_n:$,$:\bar L_n:$ operators
\begin{equation}
(L_n-a\delta_{n,0})\mid phys >=(\bar L_n-a\delta_{n,0})
\mid phys >=0, \qquad n>0.
\end{equation}
By virtue of Eq.(19) for n=0 one finds the mass of the states
\begin{equation}
 \qquad m^2={\cal P}^2=-4\pi\left\{\sum\limits_{n>0}(\alpha^\mu_{-n}
\alpha^\mu_n+\bar \alpha^\mu_{-n}\bar \alpha^\mu_n)+2a\right\}
\end{equation}
As it should be, mass of the state is determined by oscillator exitations
of $x^\mu(\tau,\sigma)$ --string only, zero modes of the
$(A^\mu_1,p^\mu_1)$ -sector do not make of contributions
into this expression.

In order to describe a spectrum of the superstring suggested in Ref.1,
it is more convenient to consider noncovariant quantization in an
appropriately chosen coordinate system. By making use of Lorentz
transformation one can consider coordinate system
where $P^\mu_y=(0,...0,1)$. ( Note that it is admissible procedure
in the canonical quantization framework, since the Lorentz
transformation is a particular example of the canonical one).
This breaks manifest $SO(1,D-1)$ covariance up to $SO(1,D-2)$ one.
In this basis Eq.(13)-(16) reduces to
\begin{equation}
 \qquad \partial_0 x^{D}=0,  \qquad \partial_0 p^{D}=0;
\end{equation}
\begin{equation}
 \qquad \partial_0 x^{\bar\mu}=-p^{\bar\mu},
 \qquad \partial_0 p^{\bar\mu}=-\partial_1\partial_1 x^{\bar\mu},
 \qquad (p^{\bar\mu}\pm\partial_1 x^{\bar\mu})^2=0;
\end{equation}
where $\mu=(\bar\mu,D)$. Thus, zero modes of the theory (1) along the
direction $P^\mu_y$ decouples from (D-1) - dimensional sector (25),
while oscillator modes along the direction $P^\mu_y$ are absent
as a consequence of the equations $(P_y\partial_1x)=(P_y\partial_1p)=0$.
As was shown in Ref.1, the same holds for the supersymmetrical case,
where equalities like $\bar\theta\Gamma^{\mu \nu}P^\nu_y\psi
=-\theta\Gamma^{\bar\mu}\psi-\bar\theta\tilde\Gamma^{\bar\mu}\bar\psi$
must be taken into account.

Let us discuss the obtained results.
The D - dimensional theory (1) can be considered as describing a pair
of strings. The sector of auxilliary variables $(A^\mu_1,p^\mu_1)$
corresponds to a string-like object which can not have oscillator
exitations. The only physical degrees of freedom of this
non-oscillating string (NO-string) are zero modes $Y^\mu$, $P^\mu_y$
which correspond to propagation of the center of mass. After
quantization, the only state of NO--string is its ground state
with the mass $m^2=-1$.

The sector of variables $(x^\mu,p^\mu)$ describes the closed string
(25),(23) which lives on (D-1) - dimensional hyperplane ortogonal
to $P^\mu_y$ - direction. (Constraints (8) relating
NO - string and the closed string mean that the last has no component
of center of mass momenta as well as of oscillator exitations
in the $P^\mu_y$ - direction, see Eqs.(17),(18)). From the mass
formulae (23), and Eq.(25) it follows that quantum states spectrum
of the theory (1) coincides with that of the (D-1) - dimensional
closed bosonic string.

In a similar fashion, spectrum of the D=11 superstring suggested
in Ref.1 coincides with that of the D=10 type IIA superstring.

\section*{Acknowledgments}
One of the authors (A.A.D) thanks N. Berkovits and J. Gates for usefull
discussions.

D.M.G. thanks Brasilian foundation CNPq for permanent support.
The work of A.A.D has been supported by the
Joint DFG--RFBR project No 96--02--00180G, and by Project INTAS--96--0308.


\begin{thebibliography}{nn}
\bibitem{1} A.A. Deriglazov, Eleven Dimensional Superstring with
New Supersymmetry and $D=10$ type IIA Green--Schwarz Superstring,
hep-th/9709025.
\bibitem{2} I. Bars, S-Theory, hep-th/9607112, hep-th/9607185.
\bibitem{3} I. Bars and C. Kounnas, A New Supersymmetry, hep-th/9612119;
String and Particle with Two Times, hep-th/9705205.
\bibitem{4} H. Nishino and E. Sezgin, SYM Equations in 10+2 Dimensions,
Phys. Lett. B 388 (1996) 569; hep-th/9607185.
\bibitem{5} I. Rudychev and E. Sezgin, Superparticles in $D>11$,
hep-th/9704057.
\bibitem{6} N. Berkovits, A Problem with the Superstring Action of
Deriglazov and Galajinsky, hep-th/9712056
\end{thebibliography}
\end{document}